\documentclass[10pt,superscriptaddress,aps,twocolumn,pra,showpacs]{revtex4-1}
\usepackage[latin1]{inputenc}
\usepackage{amsmath}
\usepackage{mathptmx}
\usepackage{graphicx}
\usepackage{subfig}

\begin{document}

\title{Dynamics of the Modulational Instability in Microresonator Frequency Combs}

\author{T. Hansson}
\email[E-mail address: ]{tobias.hansson@ing.unibs.it}
\author{D. Modotto}
\author{S. Wabnitz}
\affiliation{Dipartimento di Ingegneria dell'Informazione, Universit\`a di Brescia, via Branze 38, 25123 Brescia, Italy}

\pacs{42.60.Da, 42.65.Hw, 42.65.Ky, 42.65.Sf}

\begin{abstract}
A study is made of frequency comb generation described by the driven and damped nonlinear Schr\"odinger equation on a finite interval. It is shown that frequency comb generation can be interpreted as a modulational instability of the continuous wave pump mode, and a linear stability analysis, taking into account the cavity boundary conditions, is performed. Further, a truncated three-wave model is derived, which allows one to gain additional insight into the dynamical behaviour of the comb generation. This formalism describes the pump mode and the most unstable sideband and is found to connect the coupled mode theory with the conventional theory of modulational instability. An in-depth analysis is done of the nonlinear three-wave model. It is demonstrated that stable frequency comb states can be interpreted as attractive fixed points of a dynamical system. The possibility of soft and hard excitation states in both the normal and the anomalous dispersion regime is discussed. Investigations are made of bistable comb states, and the dependence of the final state on the way the comb has been generated. The analytical predictions are verified by means of direct comparison with numerical simulations of the full equation and the agreement is discussed.
\end{abstract}

\maketitle

\section{Introduction}

An optical frequency comb consists of a large number of highly resolved and nearly equidistant spectral lines. Frequency combs show great promise for many current and emerging applications. This includes numerous applications to spectroscopy, precision frequency metrology, optical clocks, synthesis of optical waveforms as well as channel generation for wavelength division multiplexing telecommunication systems \cite{Kippenberg2}. Many of these applications rely on creating radio-frequency beatnotes between known comb line frequencies and other unknown optical frequencies, enabling radio-frequency measurements of optical phenomena. This requires knowledge about the absolute frequency of the comb lines, which can be determined with great precision by the self-referencing method if the optical comb spans a full octave.

Frequency combs can be generated either using mode-locked femtosecond lasers or using continuous wave (CW) pumped microresonator cavities. The light is confined within a small mode volume in microresonators, which enhances the intensity dependent nonlinear interaction, thereby enabling efficient frequency conversion. Comb generation in microresonators relies fundamentally on the parametric four-wave mixing process. This nonlinear process is responsible for transferring energy from the pump mode and redistributing it among the frequency sidebands. A special degenerate case of four-wave mixing is the modulational instability, where two pump photons are annihilated to create a signal and idler pair at an equidistant frequency spacing from the pump. This mechanism is responsible for creating the primary sidebands, and is the most important frequency conversion process which occurs in microresonators.

In this paper we consider microresonator based frequency combs in the formalism of a driven and damped nonlinear Schr\"odinger (NLS) equation. Our aim is to get a better understanding of the generation of Kerr frequency combs and their long-term dynamical behaviour. The principal path to frequency comb generation is to start with an empty cavity and use the modulational instability to create sidebands from the unstable steady-state CW solution of the pump mode. Since a given comb state can be represented as a point in an infinite dimensional phase space, this can be interpreted to imply that a trajectory exists which connects the modulationally unstable CW solution with the final comb state. A comb that is generated in this manner is known as a soft excitation \cite{Matsko2}. A linear modulational instability analysis can be used to determine when the steady-state CW pump solution becomes unstable, but it cannot make any predictions about the final comb state. Moreover, stationary comb states may also exist which are isolated in phase space in the sense that their trajectories are not connected with the CW solution. These are conversely known as hard excitations. In this paper we use a finite mode truncation to limit the size of the phase space to four dimensions. This is accomplished by using a three-wave mixing model that takes into account the pump mode and the dominant sideband pair. This approximation can be justified by the fact that a substantial part of the total energy will be contained within these three modes \cite{TW}. The location of different solutions representing either stationary or breather type states can thus be determined by studying the location and stability of fixed point curves of the three-wave model.

An analysis using a similar three-mode model was previously presented in \cite{Matsko2}, c.f. also \cite{Matsko4}, where the concept of soft and hard excitations was introduced. In this paper we make a more detailed analysis, which is neither limited to the anomalous dispersion regime nor assumes a particular physical system with a fixed dispersion.

The driven and damped NLS equation and the cavity boundary condition are introduced in section II together with the normalization used throughout the rest of the article. Section III contains an analysis of the modulation instability which is used for later comparison with the three-wave model and to determine the dominant sideband pair. The truncated three-wave mixing model is derived in section IV together with a set of equations for the location of its fixed points. The influence of the comb parameters on the dynamics is considered in section V, which describes different regimes of comb generation. Section VI discusses the dependence of the final comb state on the route used to generate the comb. The final section contains conclusions and a discussion about the agreement of the model with numerical simulations.

\section{The driven and damped NLS model}

Most theoretical descriptions of microresonator frequency combs to date have been carried out using a modal expansion approach, which describes the slow evolution of the comb spectrum using coupled mode equations \cite{Matsko4,Chembo}. This approach is not very suitable for direct modelling of temporal structures and may be cumbersome to apply to broadband combs, which in the case of octave spanning combs can comprise hundreds or even thousands of resonant modes. An alternative description of microresonator frequency combs was proposed by Matsko et al. in \cite{Matsko1}, where the intracavity field is modelled in the time-domain using a mean-field driven and damped nonlinear Schr\"odinger (NLS) equation, viz.

\begin{equation}
    \tau_0\frac{\partial A}{\partial\tau} + i\frac{\beta}{2}\frac{\partial^2 A}{\partial t^2} - i\gamma|A|^2A = - \left(\alpha + \frac{T_c}{2} + i\delta_0\right)A + \sqrt{T_c}A_{in}.
    \label{eq:LLE}
\end{equation}
Where $t$ is the ordinary time that describes the temporal structure of the field inside the cavity, and $\tau$ is a slow time describing the evolution of this structure over successive round-trips. This equation is also known in the literature as the Lugiato-Lefever equation \cite{LL,CRE}, and has previously been used to model the nonlinear dynamics of dispersive fiber ring cavities \cite{HTW}, where the equation is obtained as the mean field limit of an infinite-dimension Ikeda map \cite{BD}.

In order that only resonant frequencies should contribute to the total field inside of the cavity, it is necessary to subject Eq.(\ref{eq:LLE}) to the periodic boundary condition

\begin{equation}
    A(t + \tau_0,\tau) = A(t,\tau)
    \label{eq:BC}
\end{equation}
where $\tau_0$ is the cavity round-trip time. This condition ensures frequency selectivity and is appropriate for resonators possessing a high quality factor. The spectral location of the cavity eigenfrequencies are assumed to be given by the Taylor expansion $\omega_{\mu} = \omega_0 + D_1\mu + (D_2/2)\mu^2$, c.f. \cite{Kippenberg1}, and the boundary conditions enable the frequency sampling of the resonance spectrum to be made using an equidistant frequency step with a spacing corresponding to the free-spectral-range ($D_1$). Cavity dispersion arises from the second term of Eq.(\ref{eq:LLE}) with the dispersion coefficient $\beta = -D_2(\tau_0/D_1^2)$. The generalization to higher-orders of dispersion is easily accomplished, see \cite{CRE}.

The remaining parameters in Eq.(\ref{eq:LLE}) are defined as follows: $\gamma$ is the nonlinear coefficient, $\alpha$ is the loss per cavity round-trip, $T_c$ is the coupling coefficient, $\delta_0$ is the detuning of the pump frequency and $A_{in}$ is the external pump field. In deriving Eq.(\ref{eq:LLE}) it is implicitly assumed that the cavity modes are degenerate so that light is localized in only one spatial mode. One must further neglect the frequency dependence of both absorption and coupling coefficients.

Eq.(\ref{eq:LLE}) can be simplified by normalizing it to reduce the number of parameters. It is convenient to choose the normalization so that the complete cavity loss $\bar{\alpha} = \alpha+T_c/2 = 1$ and the cavity round-trip time $\tau_0 = 2\pi$, giving

\begin{equation}
    \frac{\partial A}{\partial\tau} + i\beta\frac{\partial^2 A}{\partial t^2} - i|A|^2A = -(1+i\delta_0)A + f_0
    \label{eq:nLLE}
\end{equation}
where $\tilde{A} = \sqrt{\gamma/\bar{\alpha}}A$, $\tilde{\tau} = (\bar{\alpha}/\tau_0)\tau$, $\tilde{t} = (2\pi/\tau_0)t$, $\tilde{\beta} = (\beta/2\bar{\alpha})(2\pi/\tau_0)^2$, $\tilde{\delta_0} = \delta_0/\bar{\alpha}$ and $f_0 = (\sqrt{\gamma T_c}/\bar{\alpha}^{3/2})A_{in}$, with the tilde which has been dropped in Eq.(\ref{eq:nLLE}) referring to the new system. Eq.(\ref{eq:nLLE}) has three independent parameters, contrary to the case of the driven and damped NLS equation on the infinite line which has only two \cite{BS}. The reason for this difference is that an extra parameter appears due to the boundary condition. We have taken these parameters to be dispersion ($\beta$), detuning ($\delta_0$) and the external pump field ($f_0$). The boundary condition in Eq.(\ref{eq:BC}) is changed to $A(t + 2\pi,\tau) = A(t,\tau)$, which corresponds to a unit free-spectral-range.

\section{Modulational instability}

Eq.(\ref{eq:nLLE}) is well known to have a bistable behaviour \cite{LL}, and can have either one or three steady-state continuous wave (CW) solutions with amplitude $A_0$ for a given pump intensity $|f_0|^2$. These solutions satisfy the equation

\begin{equation}
    |f_0|^2 = |A_0|^2\left[\left(\delta_0-|A_0|^2\right)^2+1\right]
    \label{eq:Scurve}
\end{equation}
which is single valued for detunings $\delta_0 \leq \sqrt{3}$, while exhibiting bistability for detunings $\delta_0 > \sqrt{3}$. The middle, negative slope branch between $|A_0|^2 = (2\delta_0 \pm \sqrt{\delta_0^2-3})/3$ is unstable with respect to CW perturbations, i.e. perturbations at the pump frequency.

Optical frequency comb generation is usually initiated by modulational instability (MI) of the CW pump mode. This well known process is an interplay between dispersion and nonlinearity which leads to a breakup of the CW solution into a periodic pulse train. It is found in many nonlinear and dispersive wave-equations and is fundamental to the formation of solitons and solitary waves. The MI causes a transfer of energy from the pump mode to the sidebands which produces the primary sidebands of the frequency comb. However, contrary to the case of e.g. an optical fiber the MI in microresonators need not display any Fermi-Pasta-Ulam type of periodic recurrence \cite{TW}, since the pump mode can become phase locked to the driving pump. While the CW solution is always adiabatically reachable from zero initial conditions, i.e. a cavity devoid of photons, the same need not be true of different frequency comb states.

To analyze the modulational instability we make a Fourier expansion of the field $A = A(t,\tau)$, viz.

\begin{equation}
    A(t,\tau) = A_0(\tau) + \sum_k A_{k}(\tau)\exp(-i k t)
\end{equation}
where $k$ is an integer, and project the resulting equation onto a Fourier basis by multiplying with $\exp(i\mu t)/2\pi$ and integrating over $t$. This produces a system of coupled mode equations equivalent to those used in the modal expansion approach \cite{Matsko4,Chembo}. In this paper we consider the dynamics of a finite mode truncation of this infinite dimensional coupled system of equations. Specifically, a three-wave model composed of the pump mode ($A_0$) and one pair of sidebands at integer frequency $\mu$ ($A_{\mu}, A_{-\mu}$). The coupled mode equations then take the form

\begin{widetext}
\begin{align}
    & \frac{\partial A_0}{\partial\tau} = i\left(|A_0|^2 + 2|A_{-\mu}|^2 + 2|A_{\mu}|^2\right)A_0 -\left(1 + i\delta_0\right)A_0 + i2A_{-\mu}^*A_{\mu}^*A_0 + f_0, \label{eq:CM1}\\
    & \frac{\partial A_{-\mu}}{\partial\tau} = -i\beta\mu^2 A_{-\mu} + i\left(2|A_0|^2 + |A_{-\mu}|^2 + 2|A_{\mu}|^2\right)A_{-\mu} -\left(1 + i\delta_0\right)A_{-\mu} + iA_{\mu}^*A_0^2, \label{eq:CM2}\\
    & \frac{\partial A_{\mu}}{\partial\tau} = -i\beta\mu^2 A_{\mu} + i\left(2|A_0|^2 + 2|A_{-\mu}|^2 + |A_{\mu}|^2\right)A_{\mu} -\left(1 + i\delta_0\right)A_{\mu} + iA_{-\mu}^*A_0^2. \label{eq:CM3}
\end{align}
\end{widetext}

Assuming that the amplitudes of the sidebands are small, we follow the classical procedure and linearize the above system of equations around the steady-state CW solution while looking for plane-wave perturbations, satisfying $A_{\mu} = a e^{\lambda\tau}$, $A_{-\mu} = A_{\mu}^*$. The instability growth rate $\lambda(\mu)$ is then found to be given by the following dispersion relation:

\begin{equation}
    \lambda = -1 \pm \sqrt{I_0^2 - \Delta k^2}
    \label{eq:disprel}
\end{equation}
where we have introduced the pump mode intensity $I_0 = |A_0|^2$ and the wave vector mismatch $\Delta k = \beta\mu^2+2I_0-\delta_0$. It is easily seen from Eq.(\ref{eq:disprel}) that the maximum gain occurs when this mismatch is zero, i.e. for frequencies $\beta\mu^2 = \delta_0 - 2I_0$, or in the event that this equations lacks real solutions for $\mu = 0$. The growth rate for periodic perturbations is therefore generally larger, and the system is more unstable with respect to MI, than with respect to uniform CW perturbations. The maximum growth rate is found to be $\lambda_{max} = I_0 - 1$ and the steady-state solution is always stable for intensities $I_0 < 1$. The threshold condition corresponding to $\lambda = 0$, is obtained when

\begin{equation}
    \beta\mu^2 = \delta_0 - g_{\pm}
    \label{eq:threshold}
\end{equation}
where we have defined $g_{\pm} = 2I_0 \pm \sqrt{I_0^2-1}$. This equation must have real solutions, i.e. there must exist a $\mu$ corresponding to a real frequency, in order for an instability to occur. For anomalous dispersion (i.e. $\beta < 0$), we require a detuning $\delta_0 < g_+$, while for normal dispersion ($\beta >0$) the required detuning is $\delta_0 > g_-$. The solution will additionally be unstable to CW perturbations in the range $g_- < \delta_0 < g_+$.

The comb spectrum will consist of a discrete set of frequencies, due to the periodic boundary condition. This implies that the above gain spectrum must overlap with the location of at least one cavity resonance. In certain cases it may happen that the gain spectrum falls in between two resonances for a particular pump power level, in which case there is a stability window where the steady-state CW solution is stable \cite{TMOP}.

\begin{figure}[h]
\centering
\subfloat[Vs pump mode amplitude: $\delta_0 = 3, \beta = -2, \mu = 0,1,2,3$]{
  \includegraphics[width=0.5\linewidth,height=0.35\linewidth]{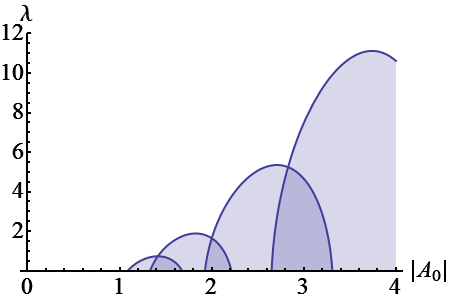}
  \label{fig:GRate_A0}}
\subfloat[Vs frequency (mode index): $\delta_0 = 3, \beta = -2, |A_0| = 1.8$]{
  \includegraphics[width=0.5\linewidth,height=0.35\linewidth]{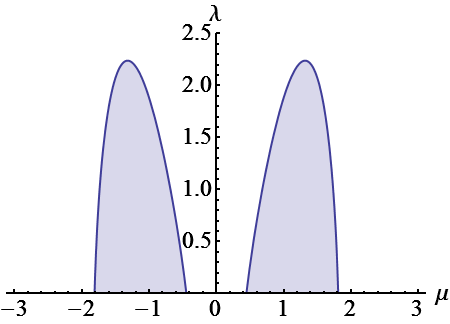}
  \label{fig:GRate_mu}}
\caption{Modulational instability growth rate for pump mode and frequency sidebands.}
\end{figure}

Fig. \ref{fig:GRate_A0} shows an example of the MI growth-rate for the first three sidebands as a function of the pump mode amplitude. The different gain curves correspond to increasing integers of $\mu$ with the first gain curve representing the CW instability, i.e. $\mu = 0$. In Fig. \ref{fig:GRate_mu}, we see a typical gain curve for a fixed pump mode intensity. Only the first sidebands is unstable in this case since the only frequencies that are allowed correspond to integer $\mu$'s.

\section{Three-wave mixing}

Although the linear MI analysis can be used to predict when the CW solution becomes unstable, and the initial growth rate of the instability, it does not provide any additional information about the subsequent dynamics or the stability of the final state. However, important insight into the dynamics can be obtained by considering finite mode truncations. The simplest such truncation is the three-wave model: consisting of the pump mode and the dominant sideband pair. This allows us to reduce the infinite dimensional phase space to one of only four dimensions. The approximation is motivated by the fact that a substantial part of the comb energy will be contained in these three modes \cite{TW}. While this procedure neglects any additional sidebands and can thus only provide an approximate description of comb generation, it is nevertheless useful for predicting the range of comb stability and the location of fixed point curves.

We consider the truncated three-wave model Eqs.(\ref{eq:CM1}-\ref{eq:CM3}) and introduce a new set of dynamical variables, viz.

\begin{align}
    & \eta = |A_0|^2/P_0, \qquad \phi = \phi_{-\mu} + \phi_{\mu} - 2\phi_0, \nonumber\\
    & P_0 = |A_{-\mu}|^2 + |A_0|^2 + |A_{\mu}|^2, \qquad \theta = \phi_{f_0} - \phi_0.
\end{align}
These correspond to the normalized pump mode intensity ($\eta$), the relative phase between pump mode and the sidebands ($\phi$), the total intensity ($P_0$), and the pump phase detuning ($\theta$), respectively. The normalized pump mode intensity $\eta$, provides a single scalar parameter that has a constant value different from one, whenever a stable comb state is reached.

We make the assumption that the sideband amplitudes are equal, i.e. $|A_{-\mu}|=|A_{\mu}|=|a|$. The difference between the amplitudes can in general be shown to be invariant. The truncated model is then found to be governed by the following set of dynamical equations:

\begin{widetext}
\begin{align}
    & \frac{\partial\eta}{\partial\tau} = -2\eta(1-\eta)\left[P_0\sin\phi - \frac{|f_0|}{\sqrt{P_0\eta}}\cos\theta\right], \\
    & \frac{\partial\phi}{\partial\tau} = (2\beta\mu^2-P_0) + 3P_0\eta - 2P_0(1-2\eta)\cos\phi - 2\frac{|f_0|}{\sqrt{P_0\eta}}\sin\theta, \\
    & \frac{\partial P_0}{\partial\tau} = -2P_0 + 2|f_0|\sqrt{P_0\eta}\cos\theta, \\
    & \frac{\partial\theta}{\partial\tau} = (\delta_0-2P_0) + P_0\eta - P_0(1-\eta)\cos\phi - \frac{|f_0|}{\sqrt{P_0\eta}}\sin\theta.
\end{align}
\end{widetext}

The limit of the system without pumping and loss was investigated in \cite{TW}. In this limit the system is conservative and reduces to two dimensions ($\eta,\phi$) so that the dynamics follows closed orbits in this plane. Seen as a projection onto the ($\eta\cos(\phi),\eta\sin(\phi)$) plane, the dynamics takes place within the unit circle, with $\eta = 1$ being a limit cycle representing the CW solution. However, dissipation is crucial to the dynamics of frequency combs, since it breaks the invariance laws of the conservative system and leads to the appearance of temporally chaotic behaviour, see \cite{HTW3}.

The stationary states are obtained by determining the fixed points of the above system, with stable equilibrium states corresponding to dynamical attractors. We consider fixed points with $\eta \neq 0,1$ and eliminating $\theta$ and $\phi$, to obtain the following two coupled equations for the absolute pump mode intensity $I_0 = |A_0|^2 = P_0\eta$ and the normalized pump mode intensity $\eta$:

\begin{widetext}
\begin{equation}
    \frac{|f_0|^2}{I_0} = \left[\left(\delta_0-I_0\right) - \frac{(1-\eta)}{\eta}g_{\pm}\right]^2 + \frac{1}{\eta^2},
    \label{eq:S1}
\end{equation}
\begin{equation}
    \beta\mu^2 = \delta_0 - g_{\pm} - \frac{3I_0}{2}\frac{(1-\eta)}{\eta}.
    \label{eq:S2}
\end{equation}
\end{widetext}

Eqs.(\ref{eq:S1}-\ref{eq:S2}) are generalizations of Eq.(\ref{eq:Scurve}) and Eq.(\ref{eq:threshold}), to which they reduce in the limit of $\eta \to 1$. It is possible to obtain a lower limit for the normalized pump mode intensity $\eta$ corresponding to the threshold intensity from Eq.(\ref{eq:S1}), which shows that

\begin{equation}
    1/|f_0| \leq \eta \leq 1, \quad 1 \leq I_0 \leq |f_0|^2\eta^2.
    \label{eq:inequals}
\end{equation}
With the limits for the total intensity of the comb being given by $P_0 = I_0/\eta$. Note that the analysis predicts that $\eta$ may become less than $1/3$, at which point the sidebands become larger than the pump mode.

Eq.(\ref{eq:S2}) gives the same result as before for the stability of the pump mode when there is no power initially in the sidebands, i.e. when $\eta = 1$. However, in the case when $\eta < 1$ and the sidebands are initially excited, there can now be additional solutions since the frequency $\mu$ is real in the anomalous dispersion regime ($\beta < 0$) whenever $\delta_0 - g_{\pm} < 3I_0(1-\eta)/2\eta$. The parameter range where solutions can exist is determined by the right hand side of the inequality, which has a minimum of zero for $\eta = 1$ and a maximum of $3I_0(|f_0|-1)/2$ for $\eta = 1/|f_0|$. Note that the parameter range for solution in the normal dispersion regime ($\beta > 0$) is smaller when the sidebands are initially excited since the solutions must now satisfy the inequality $\delta_0 - g_{\pm} > 3I_0(1-\eta)/2\eta$.

Eqs.(\ref{eq:S1}-\ref{eq:S2}) can also be combined to yield a single implicit equation for $I_0$, viz.

\begin{equation}
    \frac{3}{2}|f_0|^2 h = \left[\hat{\delta_0}-\left(\frac{5g_{\pm}}{3} - 2I_0\right)h\hat{\kappa}\right]^2 + h^2\hat{\kappa}^2
    \label{eq:implicit}
\end{equation}
where $\hat{\delta_0}=\delta_0+g_{\pm}-I_0$, $\hat{\kappa}=\kappa+2g_{\pm}-5I_0/2$, $\kappa = \beta\mu^2$ and

\begin{equation}
    h = \frac{3I_0/2}{1 + (g_{\pm} - 3I_0/2)^2}.
\end{equation}
The implicit equation is valid under the additional restriction that $\hat{\delta_0}-\hat{\kappa} \geq 3I_0/2$, corresponding to the requirement that $0 < \eta < 1$ and $I_0 > 0$.

\section{Regimes of comb generation}

% Anomalous dispersion

The three-wave model shows that comb generation in microresonators can display qualitatively different dynamics, depending on the values assumed by the free parameters and the applied initial conditions. Stable frequency comb states can, as we have argued using the phase space interpretation, be classified as either soft or hard excitations \cite{Matsko2}. A soft excitation is a comb state which can be reached in an adiabatic manner when starting from zero initial conditions. With a stable comb state, which in contrast is not adiabatically reachable from zero initial conditions, being known as a hard excitation. To reach a hard excitation state normally requires initial condition where the sidebands are already excited but they may sometimes also be reached by abrupt changes in pump intensity or detuning.

\begin{figure}[h]
\centering
\subfloat[$|f_0| = 3, \beta = -1$]{
  \includegraphics[width=0.5\linewidth]{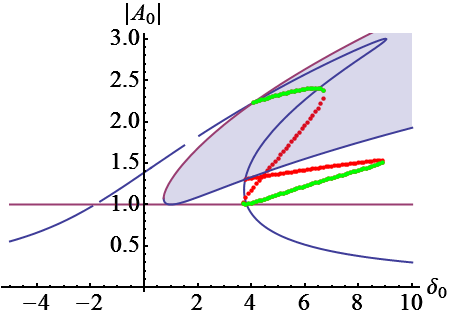}
  \label{fig:betaN1}}
\subfloat[$|f_0| = 3, \beta = -2$]{
  \includegraphics[width=0.5\linewidth]{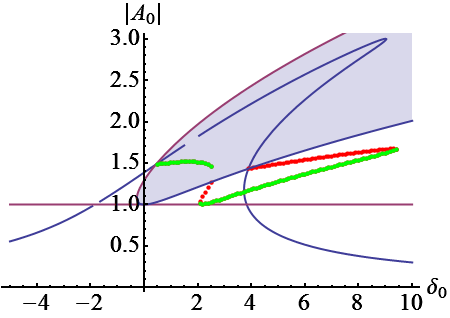}
  \label{fig:betaN2}}\\
\subfloat[$|f_0| = 3, \beta = -3$]{
  \includegraphics[width=0.5\linewidth]{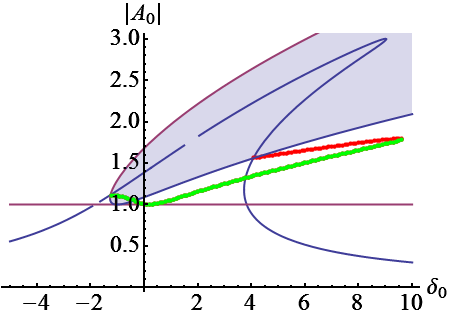}
  \label{fig:betaN3}}
\subfloat[$|f_0| = 3, \beta = -4$]{
  \includegraphics[width=0.5\linewidth]{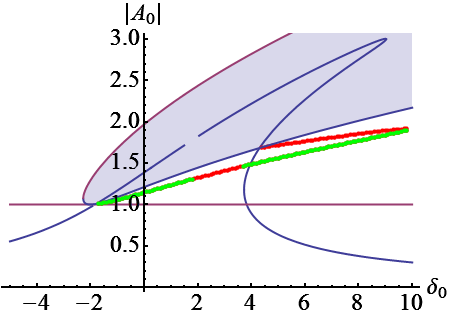}
  \label{fig:betaN4}}
\caption{Intracavity pump mode amplitude as a function of detuning for fixed external pump amplitude. The CW solution is shown by the blue curve. Green curves are stable fixed point curves while red curves are unstable. The shaded area is the region where the CW solution is modulationally unstable.}
\label{fig:parameters}
\end{figure}

The maximum growth rate for the modulational instability gives a minimum threshold intensity for the pump mode, viz. $I_0 = 1$. However, comb generation will usually not initiate at this point since the solution is part of a curve of fixed points that lies outside of the region that is modulationally unstable, c.f. \cite{Chembo}. This can be seen in Fig. \ref{fig:betaN1} which shows the amplitude of the mode pump as a function of the detuning. The blue curve shows the amplitude variation of the CW solution, while the red and green curves are the fixed point solutions of Eqs.(\ref{eq:S1}-\ref{eq:S2}), with green color denoting stable solutions and red color unstable solutions. The shaded area shows the region where the CW solutions are predicted to be modulationally unstable. The threshold intensity is indicated by the line $|A_0| = \sqrt{I_0} = 1$. The stability of the solutions is determined by considering the eigenvalues of the linearization of Eqs.(\ref{eq:S1}-\ref{eq:S2}) around the steady-state solution. The first fixed point which overlaps with a modulationally unstable CW solution of the system Eqs.(\ref{eq:S1}-\ref{eq:S2}) appears when $\eta = 1$ and $|f_0|^2 = \hat{I_0}\left[(\delta_0-\hat{I_0})+1\right]$ with $\hat{I_0} = \left[2(\delta_0-\kappa) + \sqrt{(\delta_0-\kappa)^2-3}\right]/3$. The analysis thus predicts that frequency combs may be generated whenever $\delta_0 - \kappa > \sqrt{3}$. The minimum pump intensity for this fixed point curve (upper green curve in Fig. \ref{fig:betaN1}) is obtained when the detuning $\delta_0 = \kappa + \sqrt{3}$, which implies that $\hat{I_0} = 2/\sqrt{3}$ and $|f_0|^2 = (2/\sqrt{3})\left[\left(\kappa+1/\sqrt{3}\right)^2+1\right]$, corresponding to the critical threshold power found in \cite{Chembo}. The solution at the threshold intensity $I_0 = 1$ is not predicted by the linear stability analysis, however, it can still be reached by first passing through the modulationally unstable region, and has a pump intensity that is found from Eq.(\ref{eq:implicit}) to be given by

\begin{equation}
    |f_0|^2 = \frac{5}{9}\left(\delta_0-\frac{7+8\kappa}{5}\right)^2+\frac{1}{5}\left(3+2\kappa\right)^2.
\end{equation}

We now consider frequency comb generation in the anomalous dispersion regime for a fixed pump intensity. Figure \ref{fig:betaN1} shows a case similar to that considered by Matsko et al. in \cite{Matsko2}. The upper green curve in Fig. \ref{fig:betaN1} is a fixed point curve of soft excitation solutions, which can be reached adiabatically by slowly increasing the frequency detuning of the pump until the curve of the CW solution intersects the modulationally unstable region. The lower fixed point curve of stable comb states lies outside of the MI region, and can therefore not be reached directly from the CW pump mode. However, it is still possible to generate a solution laying on this curve by slowly changing the detuning, at least for the parameter values considered in the figure. This is accomplished by first traversing the upper fixed point curve, after which the solution will jump to the lower fixed point curve. Consequently we have a case where a phase space trajectory exists which connects one stable solution set with another. The lower curve of fixed points, which is considered a hard excitation in \cite{Matsko2}, can in this way be reached in an adiabatic manner without the need for abrupt changes in either detuning or pump power.

In Fig. \ref{fig:betaN2} we consider the same parameters as in Fig. \ref{fig:betaN1}, with the exception of the dispersion which is taken to be twice as large. It is seen that the upper fixed point curve starts at a smaller detuning than before, with the discontinuity between the two curves happening outside of the bistable region. The lower fixed point curve can therefore be reached directly from the CW solution. E.g. choosing the pump detuning to be $\delta_0 = 3$ shows that the CW solution will be unable to persist, and spontaneous generation of frequency sidebands from noise induced MI will lead to a final state that lies on the lower fixed point curve. This shows that changes in the magnitude of the dispersion can change the character of the dynamics such that the trajectories of comb states which are isolated in phase space may become connected with the modulationally unstable CW solution. The distinction between soft and hard excitations must therefore be made on a case by case basis for microresonator possessing different amounts of dispersion. Fig. \ref{fig:betaN3}, where the dispersion is now three times as large as in Fig. \ref{fig:betaN2}, shows this even more clearly. Here we see that the previous two fixed point curves have merged and instead form a single solution curve, which as before is directly accessible from zero initial conditions.

Continuing to increase the dispersion reveals another interesting range of dynamics, as demonstrated in Fig. \ref{fig:betaN4}. A part of the fixed point curve is now unstable, at the same time as the corresponding CW solution, with no steady-state solution existing within the three-wave model. Numerical simulations of the three-wave model and the full Eq.(\ref{eq:nLLE}) for this case shows the solution to be periodic, corresponding to a breather type state.

% Zero dispersion

The three-wave model suggests that stable comb states could also exist for the zero dispersion case ($\kappa = 0$) as long as the system is bistable, i.e. $\delta_0 > \sqrt{3}$. These states are not predicted by the linear analysis, c.f. \cite{Chembo}, and can only be reached by hard excitations. Comb generation from MI would normally not be expected to occur for zero dispersion since MI is fundamentally an interplay between dispersion and nonlinearity which cannot take place without the presence of both effects. However, the detuning provides an extra degree of freedom which enables Eq.(\ref{eq:nLLE}) to exhibit MI in the normal dispersion regime \cite{HTW2} and also for the zero dispersion case. The theory does not predict stable soft excitation comb states close to the zero dispersion point, which may seem contradictory to the experimental observations which have been made of highly equidistant frequency combs \cite{Kippenberg3}. But one should remember that mode pulling through thermal and nonlinear effects can change the path length and the refractive index properties of the resonator so that the spectrum becomes nearly equidistant. The three-wave model thus only shows that comb generation cannot be initiated in a cold resonator through MI unless the dispersion is non-zero, while the final comb state of the warm and populated resonator may correspond to a hard excitation state. Moreover, close to the zero dispersion point it will also become necessary to include the effects of higher orders of dispersion, which will limit the attainable comb width. It should furthermore be noted that the zero dispersion case is a limit where the assumptions behind the three-wave model break down, since no particular pair of sidebands can be considered to be dominant. It is in fact easily seen that Eq.(\ref{eq:nLLE}) does not have any continuous steady-state solutions for zero dispersion, except for flat-locked CW solutions, c.f. \cite{BS}.

Regardless, numerical simulations of the driven and damped NLS equation show that ultra wideband frequency combs can indeed be generated near the zero dispersion point even if a very small anomalous dispersion is present. It is necessary to include a small amount of dispersion not only because the zero dispersion limit is unstable but also because of numerical limitations which require the comb spectrum to be finite. E.g. using parameters $\delta_0 = 3, |f_0| = 1.85, \beta = -10^{-6}$ and a starting value of $|A_0(0)| = 1.2$, produces a frequency comb consisting of nearly $40000$ resonant modes. Decreasing the dispersion further would allow the generation of even greater combs.

% Normal dispersion

\begin{figure}[ht]
  \centering
  \includegraphics[width=0.9\linewidth]{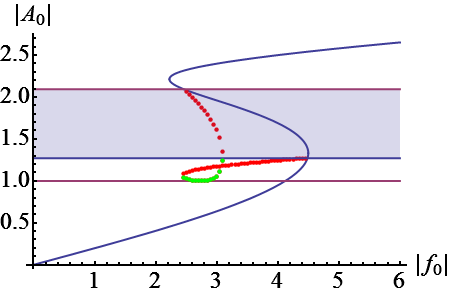}
  \caption{Hard excitation in the normal dispersion regime; intracavity pump mode amplitude as a function of external pump amplitude for fixed detuning: $\delta_0 = 5, \beta = 0.5$\\ (Same color coding scheme as Fig. \ref{fig:parameters}).}
  \label{fig:normal1}
\end{figure}

Comb generation occurs also in the normal dispersion regime. However, most of these frequency comb states are only hard excitations, c.f. \cite{Matsko3}. A typical example is found in Fig. \ref{fig:normal1}, which shows the stable fixed point curve laying inside the region exhibiting bistability of the CW solution, while outside of the MI region and not connected in any way with the CW solution curve. This fixed point curve represents an isolated solution set in phase space, whose trajectories do not connect with the state corresponding to zero initial conditions. They can therefore not be reached in an adiabatic manner and are hard excitation states. For the system to reach these solutions the initial condition need to be non-zero, which will generally require the sidebands to already be present inside the resonator. However, it may sometimes be possible to excite these states from MI of the CW solution by some abrupt change in either pump intensity or detuning: a possible route for exciting these solutions could e.g. be to rapidly change the external pump intensity in a manner so that the internal pump mode power falls into the MI region when starting from an initial state on the stable steady-state CW solution above the fixed point curve.

\begin{figure}[ht]
  \centering
  \includegraphics[width=0.9\linewidth]{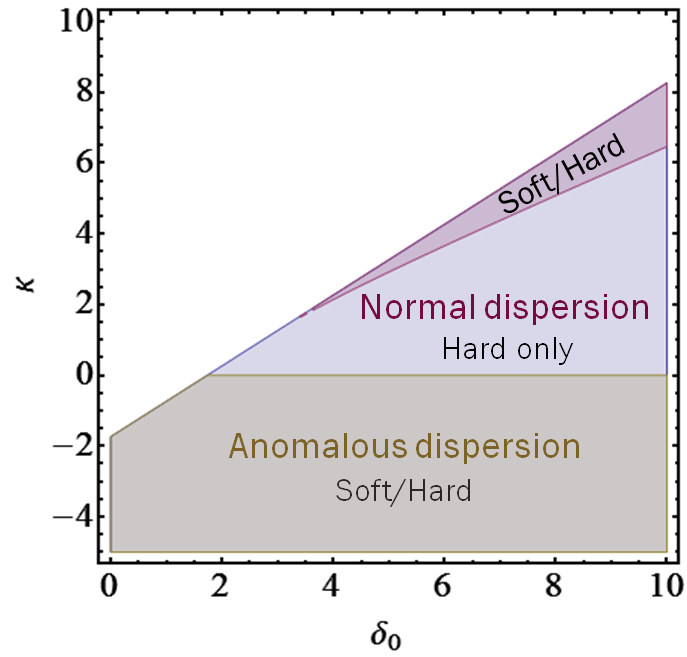}\hspace*{.4\linewidth}
  \caption{Different parameter regions where soft and hard excitations are possible.}
  \label{fig:soft}
\end{figure}

\begin{figure}[h]
\centering
\subfloat[Intracavity pump mode amplitude as a function of external pump amplitude for fixed detuning: $\delta_0 = 15, \beta = 12$]{
  \includegraphics[width=0.9\linewidth]{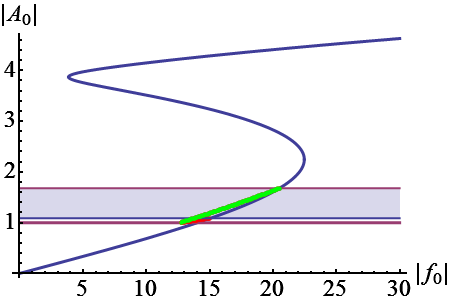}
  \label{fig:normal2}}\\
\subfloat[Numerical solution of Eq.(\ref{eq:nLLE}) for the parameters in (a): $|f_0| = 17$]{
  \includegraphics[width=\linewidth]{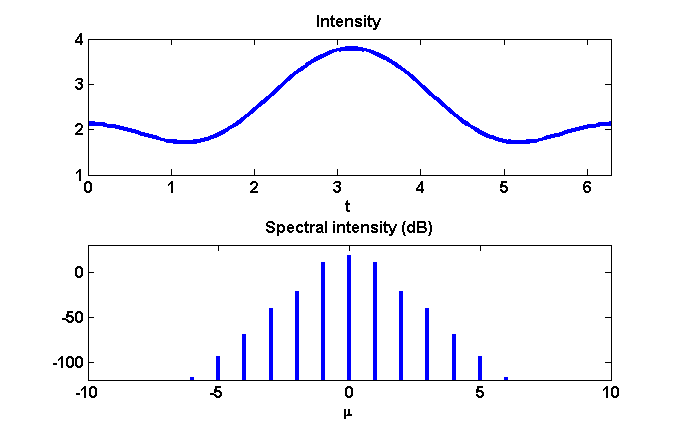}
  \label{fig:normal3}}
\caption{Example of soft excitation frequency comb generation in the normal dispersion regime.}
\end{figure}

Using the three-wave model it is found that it is also possible to generate soft excitation frequency combs in the normal dispersion regime. Such a case is demonstrated in Fig. \ref{fig:normal2}, which shows a stable fixed point curve located in the region where the CW is modulationally unstable. It can be rigourously proved that stable comb generation by means of soft excitations is possible in the normal dispersion regime within the range defined by

\begin{equation}
    \sqrt{3}+\kappa < \delta_0 < \frac{1}{2}\left(\kappa+2\sqrt{\kappa^2+4}\right)
    \label{eq:nrange}
\end{equation}
for $\delta_0 > 5/\sqrt{3}$ and $\kappa > 2/\sqrt{3}$, see Fig. \ref{fig:soft}.

However, it should be noted that the solution need not always converge to a stable comb state, even when such a state is predicted to exist by the three-wave model. The three-wave model is only capable of predicting local and not global stability. As there is often a competition between two different stable states, one being the frequency comb state and the other the steady-state CW solution, it is usually required that the MI be able to act under a sufficient long time for the system to end up in the comb state, allowing the amplitude of the sidebands to build-up. This will usually only be the case if the parameters are sufficiently large, c.f. Fig. \ref{fig:normal2}. Fig. \ref{fig:normal3} is the result of a numerical simulation of a soft excitation in the normal dispersion regime, showing the intensity and spectrum of the frequency comb. The comb was generated from noise using MI, for the parameters of Fig. \ref{fig:normal2}. A closer look at the MI gain spectrum for this case reveals that the gain for the first sideband is separated from the gain of the CW instability, by a sizable stability window which inhibits the system for jumping to the upper CW solution. The parameter range for which soft and hard excitations are possible is shown in Fig. \ref{fig:soft}. Soft excitations have been found to always be possible in the anomalous dispersion regime, and also within a part of the normal dispersion region. These soft excitations were not found by Matsko et al. in \cite{Matsko3} as they only considered a specific set of resonator parameters laying outside of the range given by Eq.(\ref{eq:nrange}).

\section{Dependence on route in comb generation}

Frequency combs can have either one or two stable curves of fixed points for a particular set of parameters. The resonator thus exhibits bistability and hysteresis not only for the CW solution but also for comb states when the sidebands are excited. It therefore becomes important to consider the way that frequency combs are generated, since variations in the route may produce different final states.

\begin{figure}[ht]
\centering
\subfloat[Bistable behaviour of comb states; intracavity pump mode amplitude as a function of external pump amplitude for fixed detuning: $\delta_0 = 5, \beta = -2.5$]{
  \includegraphics[width=0.9\linewidth]{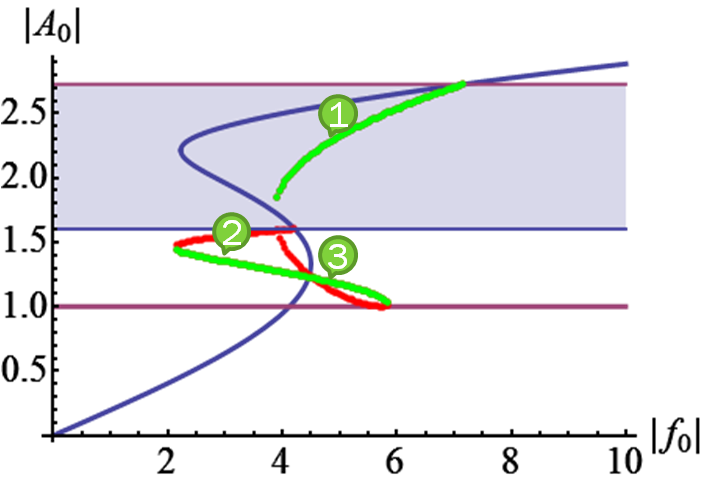}
  \label{fig:route}}\\
\subfloat[Numerical solution of Eq.(\ref{eq:nLLE}) for the parameters in (a) showing the spectral evolution for varying pump intensity. Logarithmic intensity scale with yellow color denoting highest intensity.]{
  \includegraphics[width=0.9\linewidth]{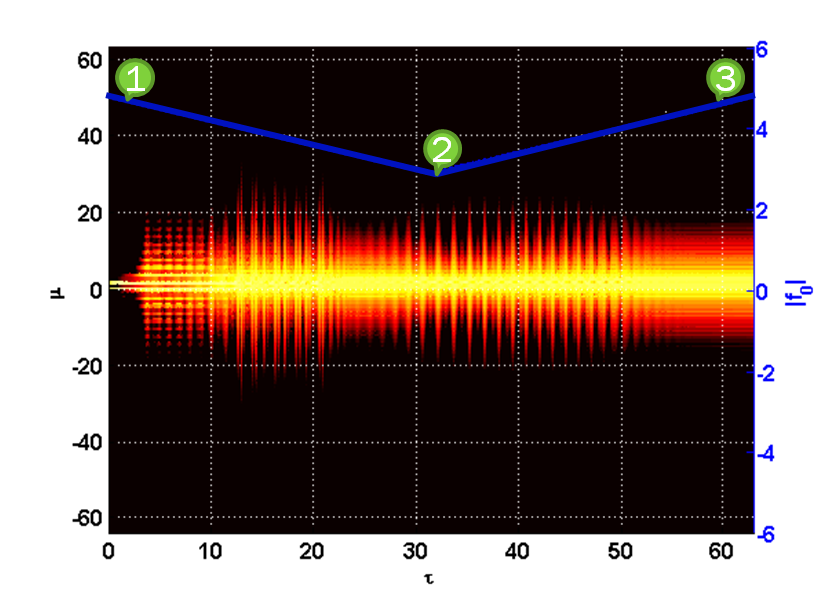}
  \label{fig:route2}}
\caption{Simulation results showing bistable behaviour and dependence of route in frequency comb generation.}
\end{figure}

Fig. \ref{fig:route} shows an example where two different fixed point curves exist for the same set of comb parameters. We consider a case where the field inside the resonator is initially zero and the pump intensity is set above the threshold intensity for bistability of the CW solution. The pump mode amplitude will then first rise to a level corresponding to the CW solution on the upper curve, but since this solution is in the MI region, it will be unstable and the system will instead approach the stable upper fixed point curve (point 1). However, if the external pump intensity is later slowly reduced beyond the point where the upper curve ends, then the system will jump down to the lower fixed point curve (point 2). If the pump intensity is now slowly increased again to its original value, then we find that the system will stay on the lower curve (point 3). A similar procedure involving the detuning would also allow us to reach different comb state corresponding to identical comb parameters in an adiabatic manner.

A numerical simulation showing this is found in Fig. \ref{fig:route2}, which is a plot of the spectral intensity as a function of the evolution time. The figure was obtained by solving Eq.(\ref{eq:nLLE}) with a time varying pump intensity $|f_0|^2$ as illustrated by the blue line in the figure (right hand scale). The frequency comb is clearly seen to display qualitatively different dynamics for the first half of the figure as compared to the second half. E.g. even though the external parameters are the nearly same, we find the comb state at $\tau \simeq 60$ to correspond to a stable soliton state with a stationary intensity profile while the comb at $\tau \simeq 5$ is found to be a temporally chaotic, breather type state. Clearly these solutions correspond to different fixed points as predicted by the three-wave model. The dynamics is also seen to abruptly change when the pump intensity is reduced below the endpoint of the upper fixed point curve, again in good agreement with the model.

\section{Conclusions}

While the three-wave model can only give an approximation of the full comb dynamics, it often produces remarkably good agreement. The three-wave model is the lowest order non-trivial finite mode truncation that can be made. The inclusion of additional modes could provide an even better approximation, but requires a more involved analysis since the phase space dimensionality grows quickly. Comparisons between numerical simulations of both of the three-wave mixing model and the full driven and damped NLS Eq.(\ref{eq:nLLE}), have shown that the three-wave model is useful in finding the location of frequency comb states as well as helping to predict where comb generation may be stable. Unfortunately, it is not possible to use the model to predict absolute comb stability since higher order sidebands are neglected. Pump mode and sidebands amplitudes are also not always accurate, especially for large frequency sidebands.

The three-wave model could be useful as a tool, together with numerical simulations of Eq.(\ref{eq:nLLE}), to help design microresonator devices capable of generating specific comb states. It could e.g. be applied to finding specific routes which generates soliton trains or octave spanning frequency combs. However, it should be remembered that if mode pulling due to thermal and nonlinear effects is significant, it may be necessary to also consider temporal changes in the dispersion, in order to accurately model the complete excitation dynamics observed under experimental conditions.

In this article we have made a study of the dynamics of the modulational instability of microresonator based frequency combs in the context of a formalism provided by the driven and damped NLS equation. We have demonstrated that the primary path to comb generation relies on the modulational instability of the CW pump mode, although other comb excitation routes are also possible and indeed necessary to reach certain comb states. A linear stability analysis has been made, which has taken into account the proper boundary conditions of frequency selective, high Q-factor, microresonators. Additionally, we have derived a truncated three-wave mixing model which describes both the dynamics and long-term behaviour of different frequency comb states in a reduced, four dimensional, phase space. The fixed points of the dynamical system have been identified, with stable states corresponding to either stationary or chaotic frequency combs. A discussion of different regimes of frequency comb generation has been made. This discussion has highlighted the role of the comb parameters in determining the excitation dynamics and shown that not only the sign of the dispersion but also its magnitude is crucial in determining the dynamical behaviour and different excitation routes. We have found that soft excitation comb generation is possible even in the normal dispersion regime, and derived a range which shows where such excitations can occur. Finally we have considered the dependence on route in comb generation, and the fact that resonators may exhibit bistable behaviour not only for the CW solution but also for different frequency comb states.

\section*{Acknowledgements}

This research was funded by Fondazione Cariplo, grant no. 2011-0395.

\end{document}